\documentclass[conference]{IEEEtran}
\IEEEoverridecommandlockouts
\usepackage{cite}
\usepackage{amsmath,amssymb,amsfonts}
\usepackage{algorithmic}
\usepackage{graphicx}
\usepackage{textcomp}
\usepackage{xcolor}
\usepackage{bbding}
\usepackage{makecell}
\usepackage{algorithm}  
\usepackage{color}
\usepackage{multirow}
\usepackage{array}
\usepackage{booktabs}
\usepackage{tabularx,booktabs}
\usepackage{multicol}
\usepackage{cleveref}
\usepackage{balance}
\usepackage{lastpage}
\usepackage{tabulary}
\usepackage{subfigure}
\usepackage{bbm}
\usepackage{ulem}
\usepackage{mathrsfs}
\usepackage{tablefootnote}
\usepackage[cal=boondoxo]{mathalfa}
\usepackage{fancyhdr} % 添加页眉页脚
\def\BibTeX{{\rm B\kern-.05em{\sc i\kern-.025em b}\kern-.08em
    T\kern-.1667em\lower.7ex\hbox{E}\kern-.125emX}}

\def\BibTeX{{\rm B\kern-.05em{\sc i\kern-.025em b}\kern-.08em
    T\kern-.1667em\lower.7ex\hbox{E}\kern-.125emX}}

\renewenvironment{IEEEbiography}[1]
  {\IEEEbiographynophoto{#1}}
  {\endIEEEbiographynophoto}
  
\begin{document}

\title{ Knowledge Transfer and Reuse: A Case Study of AI-enabled Resource Management in RAN Slicing \\
\thanks{

H. Zhou and M. Erol-Kantarci are with the School of Electrical Engineering and Computer Science, University of Ottawa, Ottawa, ON K1N 6N5, Canada. (emails:\{hzhou098, melike.erolkantarci\}@uottawa.ca).

H. V. Poor is with the Department of Electrical and Computer Engineering,
Princeton University, Princeton, NJ 08544 USA (e-mail: poor@princeton.edu).}}
\author{\IEEEauthorblockN{Hao Zhou, \IEEEmembership{Graduate Student Member, IEEE}, Melike Erol-Kantarci, \IEEEmembership{Senior Member, IEEE}, \\  and Vincent Poor, \IEEEmembership{Life Fellow, IEEE}}}
\maketitle
\thispagestyle{fancy}            %更改plain状态，首页格式设为fancy
\chead{This paper has been accepted by IEEE Wireless Communications Magazine }                     %清除以前的命令
%\rhead{righthead}                %页眉右侧内容
%\lfoot{leftfoot}                 %页脚左侧内容
%\cfoot{\quad}                    %没找到清除页码的，直接用空格覆盖住页脚中间页码
\renewcommand{\headrulewidth}{0pt}      %把页眉线的宽度设为零，即去掉页眉线
\pagestyle{plain}                %首页后的章节格式设置为空

%TC:ignore
\begin{abstract}  
An efficient resource management scheme is critical to enable network slicing in 5G networks and in envisioned 6G networks, and artificial intelligence (AI) techniques offer promising solutions. Considering the rapidly emerging new machine learning techniques, such as graph learning, federated learning, and transfer learning, a timely survey is needed to provide an overview of resource management and network slicing techniques of AI-enabled wireless networks. 
This article provides such a survey along with an application of knowledge transfer in radio access network (RAN) slicing. In particular, we first provide some background on resource management and network slicing, and review relevant state-of-the-art AI and machine learning (ML) techniques and their applications. Then, we introduce our AI-enabled knowledge transfer and reuse-based resource management (AKRM) scheme, where we apply transfer learning to improve system performance. Compared with most existing works, which focus on the training of standalone agents from scratch, the main difference of AKRM lies in its knowledge transfer and reuse capability between different tasks. Our paper aims to be a roadmap for researchers to use knowledge transfer schemes in AI-enabled wireless networks, and we provide a case study over the resource allocation problem in RAN slicing. 
\end{abstract}

%TC:endignore

\section{Introduction}
\label{s1}

Network slicing is a fundamental concept for 5G networks and envisioned 6G networks that enables multiple services and applications with heterogeneous demands. With software-defined networks and network function virtualization techniques, various slices can be defined over the same physical network devices for flexibility and scalability \cite{b1}. However, network slicing also leads to significant complexity for resource management. 
Compared with core network slicing, radio access network (RAN) slicing requires more efficient resource management to adequately utilize the limited bandwidth resources. In addition, user types are not limited to enhanced Mobile Broad Band (eMBB), Ultra Reliable Low Latency Communications (URLLC), and massive Machine Type Communications (mMTC). An enterprise slice with various demands can be dynamically created, and these demands need to be fulfilled in a much shorter time scale than the core. Moreover, dynamic traffic patterns, increasing device numbers, and emerging new services also contribute to the technical challenges of RAN slicing.  
Such complexity of RAN slicing prevents the applications of conventional optimization methods, especially considering a dynamic wireless environment. To this end, artificial intelligence (AI) and machine learning (ML) offer promising opportunities \cite{b2}. For instance, convex optimization must design dedicated optimization models for each problem, while reinforcement learning (RL) can significantly reduce the optimization complexity by transforming problems into a unified Markov decision process (MDP).

\begin{figure*}[!t]
\centering
\vspace{-7pt}
\includegraphics[width=16cm,height=9.1cm]{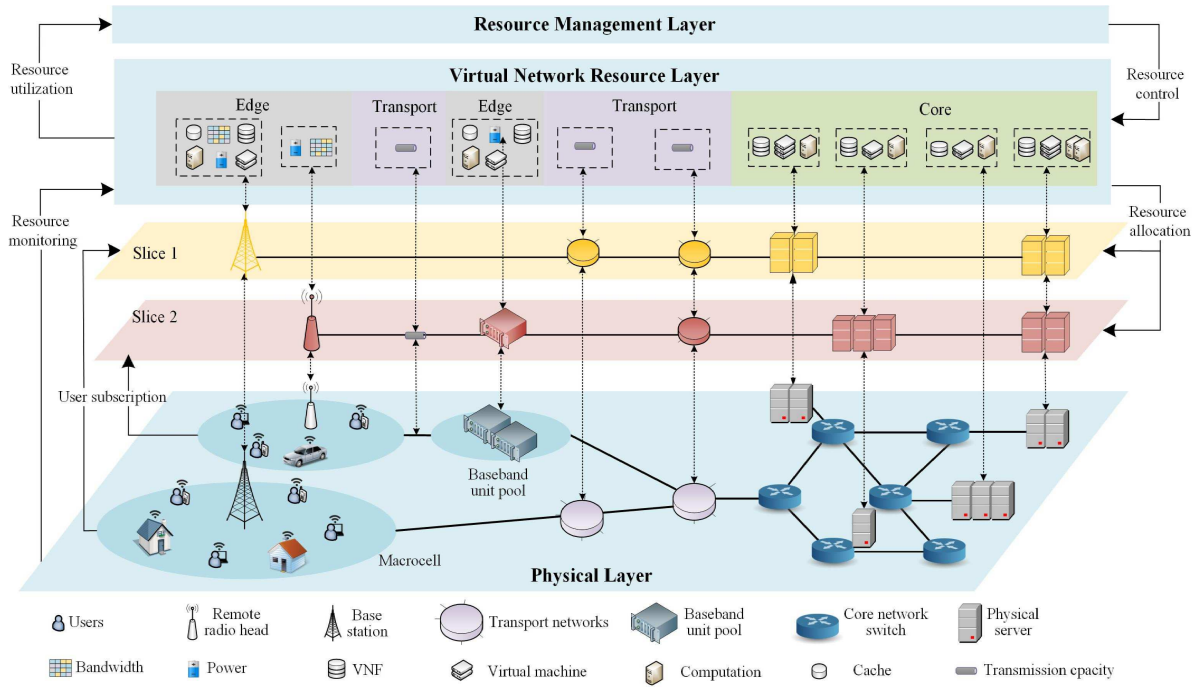}
\caption{Overall architecture of the defined resource management scheme.}
\label{fig1}
\vspace{-14pt}
\end{figure*}

AI-based network management has attracted interest from both academia and industry. The Management and Orchestration (MANO) is the most notable framework defined by European Telecommunications Standards Institute (ETSI)\cite{b3}. Similarly, the Third Generation Partnership Project (3GPP) also defined the network data analytics function and management data analytics function to incorporate AI into resource management\cite{b4}. These ingeniously designed schemes, nevertheless, are still ongoing, and these frameworks mainly focus on system-level definitions of key elements. 

Meanwhile, AI-enabled wireless networks have been extensively investigated by academia. For example, \cite{b5} summarizes the applications for employing AI in cellular networks, and identifies the challenges and roadmaps for AI-enabled 5G and 6G networks. A thorough survey is presented in \cite{b6} on recent advances and future challenges of applying AI to wireless networks, including algorithms, applications, standards and so on. However, some newly emerging techniques, such as graph learning and federated learning, are not included in these works, which calls for a timely survey on the latest ML methods as applied to problems in wireless communications. 

On the other hand, numerous schemes have been proposed to make the most of AI for network management. For instance, AI is considered a built-in architectural feature in \cite{b7} that enables resource elasticity, and it includes different use cases that utilize AI for elastic management. 
In these studies, plenty of samples are generally required to train their algorithms, such as deep Q-learning (DQN), and the long training time may hamper the system efficiency and their applicability in practice. Moreover, in these schemes, the costly training needs to be repeated for each arriving new problem, and this low generalization capability inspires us to find a more efficient architecture to utilize AI/ML.

Motivated by these challenges, in this work, we first provide some background on resource management and summarize various network resources, ranging from the RAN to the core network. Then, we review state-of-the-art AI/ML algorithms for resource management such as graph learning, federated learning, transfer learning, etc. Furthermore, we introduce an AI-enabled knowledge transfer and reuse-based resource management (AKRM) scheme. AKRM is designed to transfer the knowledge of source tasks and reuse them for target tasks. In particular, we apply transfer learning to digest the experience of expert agents, which will be further utilized by learners to reduce the algorithm training efforts. Compared with our former work \cite{b9}, here we give a more systematic roadmap of using transfer learning that includes multiple experts and learners simultaneously. 

The contributions of this work are two-fold: first, we present an up-to-date survey of AI/ML algorithms and their applications on resource management; second, we define an AKRM scheme that enables knowledge transfer and reuse for network management problems. In addition, we introduce transfer deep reinforcement learning (TDRL) and transfer learning-based neural networks as examples to better explain how to reuse the prior knowledge. Finally, we provide a case study of TDRL-based resource management in RAN slicing, which proves the superiority of AI-enabled knowledge transfer. 

The remainder of this article is organized as follows. Section \ref{s2} provides background on resource management in 5G, and Section \ref{s3} surveys relevant AI techniques. Section \ref{s4} defines the AKRM scheme, and Section \ref{s4-2} introduces transfer learning-based knowledge reuse. Section \ref{s5} shows the case study, and Section \ref{s6} concludes this paper.  

\begin{table*}[!t]
\caption{Summary of network resources in 5G networks.}
\centering
\vspace{0pt}
\renewcommand\arraystretch{1.4}
\begin{tabular}{|m{1cm}<{\centering}|m{1.4cm}<{\centering}|m{6cm}<{\centering}|m{2.5cm}<{\centering}|m{5cm}<{\centering}|}
\hline
Domains & Resources & Features  & Applications & Typical management objectives \\
\hline
\multirow{12}*{\makecell{RAN \\ and \\edge}} & Radio & Spectrum requires careful allocation, such that transmission requirements of various slices are satisfied using this very scarce resource.  & Spectrum allocation & Maximizing throughput or minimizing the delay; improving spectrum efficiency.  \\
\cline{2-5}
~& Power & Power is a key resource in RAN. Power control can mitigate interference and improve energy efficiency for network devices.  & Power allocation for interference management & Minimizing interference and energy cost; maximizing network throughput.\\
\cline{2-5}
~& Cache  & Edge caching can significantly reduce the network delay by storing popular contents in the edge, and caching capacity is a limited resource. &  Content replacement and delivery in network edge. & Maximizing hit ratio in a cache for better network performance (lower delay) or economic metrics. \\
\cline{2-5}
~& Computation & Computation capacity is a key resource in the edge, where computationally intensive tasks from user devices (UEs) or IoT devices can be processed in the network edge instead of the cloud. & MEC offloading between UEs, MEC server and cloud. & Minimizing processing delay or computation cost. \\
\cline{2-5}
~& VNFs, VMs, containers & VMs or containers are required when VNFs are deployed in the mobile edge, and thus they become important resources to run VNFs. & VMs and VNF placement. & Minimizing response time, energy cost and hardware usage. \\
\hline
\multirow{3}*{\makecell{Transport}} & Fronthaul  & Connections between remote radio heads and baseband unit pool in the edge cloud RAN. & \multicolumn{2}{|c|}{\multirow{2}*{ \shortstack{The connection capacity is considered as \\ critical transmission constraints. Pricing models \\can be applied to maximize the profit by managing \\the transmission capacity allocation. }}} \\
\cline{2-3}
 & Backhaul & Connections from RAN to the core network, or from edge (base stations or baseband unit pools) to the core network.  & \multicolumn{2}{|c|}{~} \\
\hline
\multirow{4}*{\makecell{Core}} & VMs and VNFs & Compared with mobile edge, core network is capable of instantiating a higher number of VMs to implement VNFs. The number of deployed VMs determines the service capability of a single network node.  & VMs and VNF placement. &  Minimizing cost or energy consumption; guaranteeing network performance and reliability.\\
\cline{2-5}
 ~& Computation & In the core network, available computation resources of physical hosts in network nodes should be allocated to VNFs to process tasks. & CPU allocation; work load assignment. & Minimizing computation delay or cost.  \\
\hline
\end{tabular}
\label{tab1}
\vspace{-5pt}
\end{table*}

\section{ Background on Resource Management for Network Slicing in 5G}
\label{s2}
Network slicing presents demanding requirements for resource management, especially at the edge and radio domains. Tight network resources are allocated concurrently between multiple slices in a very short timescale, to satisfy different service level agreements. Although managing one single resource has been widely studied in the literature, the evolving network architecture requires the holistic management of multiple network resources. For example, in the multiple-access edge computing (MEC)-enabled 5G RAN, the computation offloading and bandwidth allocation must be jointly considered to guarantee the network performance. Furthermore, end-to-end network slicing calls for cross-domain resource allocation that includes RAN, edge, transmission, and core networks \cite{b11-1}. Thus a unified architecture is needed to include all the cross-domain resources in a systematic view.  

Fig.\ref{fig1} shows our defined resource management architecture, including the physical layer, virtual network resource layer, and management layer. First, the physical layer consists of physical network devices that can be split into different network slices. For instance, slices 1 and 2 in Fig.\ref{fig1} may require multiple network resources simultaneously, and one single physical device can be shared by both slices. Then, by monitoring the status of the physical layer, we define a virtual network resource layer to aggregate the resources of different network elements. For example, the baseband unit pool in Fig.\ref{fig1} includes power, cache, computation resources, while the remote radio head mainly includes bandwidth and power as manageable resources. Finally, the management layer works as a control plane to manage virtual resources, and these resources will be allocated to slices to fulfill their requirements. The management layer details will be introduced in Section \ref{s4}. 

Meanwhile, we summarize diverse network resources in Table \ref{tab1}, including the features, applications, and management objectives. Note that the computation resource allocation is involved in both edge and core networks. The network edge devices usually have limited computation capacity, which makes the computation task offloading a critical concept. On the contrary, the core network has more abundant computation capacities, and more computationally intense tasks without real-time requirements can be processed. Virtual machines (VMs) can also be deployed in both edge and core networks to implement virtual network functions (VNFs), but the edge side has limited hardware resources to support the VMs or containers. Table \ref{tab1} implies that resource management of RAN and edge is more complicated than transport and core networks due to limited available resources, more dynamic conditions, and the need for actions on a shorter timescale.

To support the applications shown in Table \ref{tab1}, an efficient control system is expected to process a wide variety of resource management tasks for network slicing, and the flourishing AI/ML methods offer promising solutions, which will be introduced in the next section.    

\begin{table*}[!t]
\caption{Summary of AI techniques for 5G resource management.}
\centering
\renewcommand\arraystretch{1.4}
\begin{tabular}{|m{1.5cm}<{\centering}|m{1.7cm}<{\centering}|m{5.5cm}<{\centering}|m{4cm}<{\centering}|m{3cm}<{\centering}|}
\hline
Learning methods & Typical algorithms & Main features & Difficulties & Typical applications\\
\hline
\multirow{1}*{\makecell{Supervised\\ learning}} & Artificial neural network, RNN, LSTM, support vector machine, etc. & Training the algorithm to best map inputs to outputs in a dataset, and thus labeled data is required for training. Note that each algorithm has its features, e.g., LSTM can better handle the long-term dependencies of data. & Algorithm training relies heavily on fine-grained datasets, which may be inaccessible in practice. Network training can be time-consuming due to hyperparameter tunings such as learning rate and the number of hidden layers or units.   & Traffic load prediction and classification for slices.\\
\hline
Unsupervised learning  & K-means, DBSCAN & Discovering hidden patterns or similarities in unlabeled datasets.  & High computation complexity; noisy results. & User clustering to create new slices. \\
\hline
\multirow{11}*{\makecell{Reinforcement \\ learning}}  & Q-learning & Agent interacts with an environment to maximize the expected reward. A Q-table is used to record state-action values in Q-learning. & Requiring long convergence time, especially for problems with large state-action space.  &\multirow{11}*{\makecell{RL is the most generally \\ applied AI technique for \\ the resource management \\ and optimization \\ wireless networks, \\ e.g., minimizing \\ network delay or energy \\ consumption, maximizing \\ network throughput or \\ revenue. For example, \\ multi-agent reinforcement\\ learning may be used \\ for resource allocation \\ of slices. }}\\
\cline{2-4}
& Actor-critic learning & The actor is considered as a policy structure to select actions, and the critic will estimate the value function for the actions of the actor. & Long convergence time; instability caused by the interplay between actor and critic. & \\
\cline{2-4}
& Deep reinforcement learning & Combining neural networks with RL framework to predict state-action values instead of using Q-tables. DRL can be applied to address the large state-action space problem. & \multirow{3}*{\makecell{Time-consuming network training; \\ tedious hyperparameter tuning; \\ network training stability;\\ low sample efficiency.}}  &\\
\cline{2-3}
& Double deep Q-learning & Decoupling the action selection and evaluation of deep Q-learning to prevent overestimation and provide a better Q-value estimation.  & &\\
\cline{2-4}
& Multi-agent reinforcement learning & Each agent implements the RL or DRL independently to achieve its goal or optimize an overall objective. & The coordination mechanism of multiple agents requires dedicated design.  &\\
\hline 
Federated learning & Federated deep learning, federated DRL & Training models by distributed datasets of multiple learners, then feeding back global model parameters to local learners for their use. Private and sensitive data can be well protected with federated learning. & Communication overhead for updating local models; systems heterogeneity caused by diverse local devices such as different storage and computation capacity.  & Each slice can train a local model, and the central controller produces a global model for coordination. \\
\hline
Graph learning & Graph neural networks, graph convolutional networks & Graph learning maps the feature of a graph to the vectors without projecting the graph into a low dimensional space, which can be used for classification, link prediction, and matching. & Scalability issue; generative graph learning; dynamic graph.  & GNN-based digital twin for network slicing. \\
\hline
\multirow{4}*{\makecell{Transfer \\learning}}  &Supervised transfer learning & Improving generalization capability by reusing the pre-trained model such as neural networks of source tasks to related target tasks.  & Task mapping function definition; negative knowledge transfer. & Classification and prediction for resource management and RAN slicing.  \\
\cline{2-5}
~ & Transfer reinforcement learning & Utilizing knowledge of experts to improve learner's performance within the context of MDP, which aims at faster convergence and higher average reward. & The knowledge transfer function requires dedicated design. &  Optimization of network resource management and RAN slicing with fewer training samples and faster convergence.  \\
\hline
\end{tabular}
\label{tab2}
\vspace{-5pt}
\end{table*}

\section{Overview of AI/ML Techniques for Resource Management in Wireless Networks}
\label{s3}

This section will briefly review the latest AI/ML techniques for resource management applications\footnote{Note that the main goal of this work is not to review all the AI/ML techniques on wireless networks but serves as a compressed taxonomy for new techniques\cite{b5}.}. Table \ref{tab2} summarizes the algorithms, main features, difficulties, and typical applications of each learning method.

\subsection{Supervised and Unsupervised Learning}
Supervised learning algorithms are designed to map input data to labeled output data for classification or prediction. Well-known algorithms and neural networks include artificial neural networks, recurrent neural networks (RNN), long short-term memory (LSTM) networks, support vector machines, decision trees, and so on. Supervised learning can be used to predict the network load of each slice, which will improve the resource allocation between slices. However, supervised learning suffers from the fact that network training is too much dependent on hyperparameter tuning and the difficulty in obtaining labeled data. Insufficient data may prevent the application of supervised learning on network slicing.

By contrast, unsupervised learning intends to analyze and cluster unlabeled data. Well-known approaches include k-means and Density-Based Spatial Clustering of Applications with Noise (DBSCAN) algorithm. The deep belief network can also be used for unsupervised learning to extract the features of unlabeled data. These techniques discover hidden similarities and differences of unlabeled data. Nevertheless, the long training time and high computational complexity are potential issues for unsupervised learning. Meanwhile, the output results are hard to validate due to the unlabeled data set. Unsupervised learning can cluster users with various service requirements as different slices, creating new slices based on real-time user demands.

\subsection{Reinforcement Learning}
RL has been widely used for resource management optimizations such as physical resource block allocation, power allocation, and so on. RL algorithms include Q-learning, actor-critic learning, deep reinforcement learning (DRL), etc \footnote{There are many versions of RL techniques in the literature. Summarizing all is out of the scope of this paper. We provide examples of the most widely used algorithms.}. In RL, the agent aims to maximize the long-term expected reward within the MDP. Q-learning is the most widely used algorithm due to its simplicity, however it suffers from a long convergence time issue, especially for problems with large state-action space.
Meanwhile, actor-critic learning defines an actor for action selection and a critic for action evaluation, which uses the value function as a baseline for policy gradients. DRL has been proposed to overcome the slow convergence problem of Q-learning and also the challenge of storing a large Q-table. Instead of deploying huge Q-tables, the state-action values in DRL are predicted by neural networks, and thus it can better handle large state-action space problems. 
%Furthermore, double deep Q-learning is designed to prevent the over-estimation issue of DQN, in which the main network selects actions and the target network evaluates actions. 

%By transforming diverse optimization problems into unified MDPs, RL methods can avoid the possible complexity of defining a dedicated optimization model. 
RL methods have been used in many resource management and network slicing problems. 
For example, DRL can overcome the huge state-action space issue caused by the increasing number of network slices and user devices.  
Different slices with heterogeneous requirements can also be considered independent agents to make autonomous decisions, which can be solved by multi-agent reinforcement learning techniques.
Nevertheless, RL algorithms demand plenty of samples for training, and the trained algorithm can only handle specific problems with poor generalization capability. Therefore, the slow convergence may hamper the resource management efficiency of network slicing.

\subsection{Federated Learning}
Federated learning is one of the state-of-the-art ML methods applied in wireless networks. Federated learning algorithms are trained across multiple decentralized local datasets without sharing training data, where only the local model parameters are sent back to a central server for global model aggregation. Then, well-trained global model parameters are fed back to local learners to update local models. As an example in wireless communications, each slice can train a federated learning model using local data, and then the central controller can form a global model and send it back to slices as in \cite{b12}. Federated learning addresses data privacy and security issues, since each learner can keep its training data on board. However, the learners still need to exchange the parameters of the trained models. As such, the communication overhead becomes a critical issue for federated learning, especially considering that multiple slices may need to exchange parameters with the central controller simultaneously. Meanwhile, considering various storage, computation, and communication capabilities of local devices, system heterogeneity can be another challenge for applying federated learning on wireless networks.

\subsection{Graph Learning}
Graphs are widely used to represent the network architecture, and graph learning refers to ML algorithms on graphs. In particular, graph learning is designed to extract the main features of graphs, such as node and edge connections, to vectors with the same dimensions in the embedding space. For instance, graph neural networks aim to learn a state embedding that contains the neighborhood information of each node, and the graph data is directly mapped to the output of the neural network without projecting the graph into a low-dimensional space. Graph learning has been successfully applied in node classification, graph classification, link predictions, and so on, which shows superior performance than other existing approaches. Considering the graph nature of wireless networks, graph learning shows significant potential for resource management and network slicing. For instance, \cite{b13} proposed a graph neural network (GNN)-based digital twin for network slicing, in which the GNN is trained to simulate the behaviors of slices. Meanwhile, graph learning still faces many challenges, e.g., dynamic graph, scalability issues, and generative graph learning. For example, users with high mobility may lead to dynamic graph structures, but graph learning architectures cannot change adaptively if edges and nodes appear or disappear frequently.         

\subsection{Transfer learning}
Although various ML methods have been introduced before, it is worth noting that: i) most algorithms require a large number of training samples to explore the target task;  ii) these algorithms are designed for specific tasks with very limited generalization capability. As a result, even though similar tasks have been completed before, the algorithms still need to be retrained for new problems. To this end, transfer learning aims to reuse the knowledge and models of existing tasks. Specifically, a model developed for one task is reused as the starting point for other related tasks. Indeed, humans can apply their knowledge from previous work to solve new works more rapidly. This reduces the need for a large number of training samples, which is a common issue in ML. As an example, the central controller can transfer the pre-trained neural network models to slices to initialize local model training, achieving a faster convergence\cite{b10}. 
Despite the potential advantages, the knowledge may exist in different forms in ML algorithms, and knowledge transfer methods vary between different tasks. %For instance, Q-values, learning rate, and action selections can all be considered knowledge. 
Meanwhile, since each slice may have different requirements, the mapping functions require sophisticated designs to make prior knowledge digestible for learners \cite{b9}.

\section{AI-enabled Knowledge Transfer and Reuse-based Resource Management}
\label{s4}

\subsection{AI-driven Resource Control Module }
\begin{figure*}[!t]
\centering
\vspace{-10pt}
\includegraphics[width=17.5cm,height=10cm]{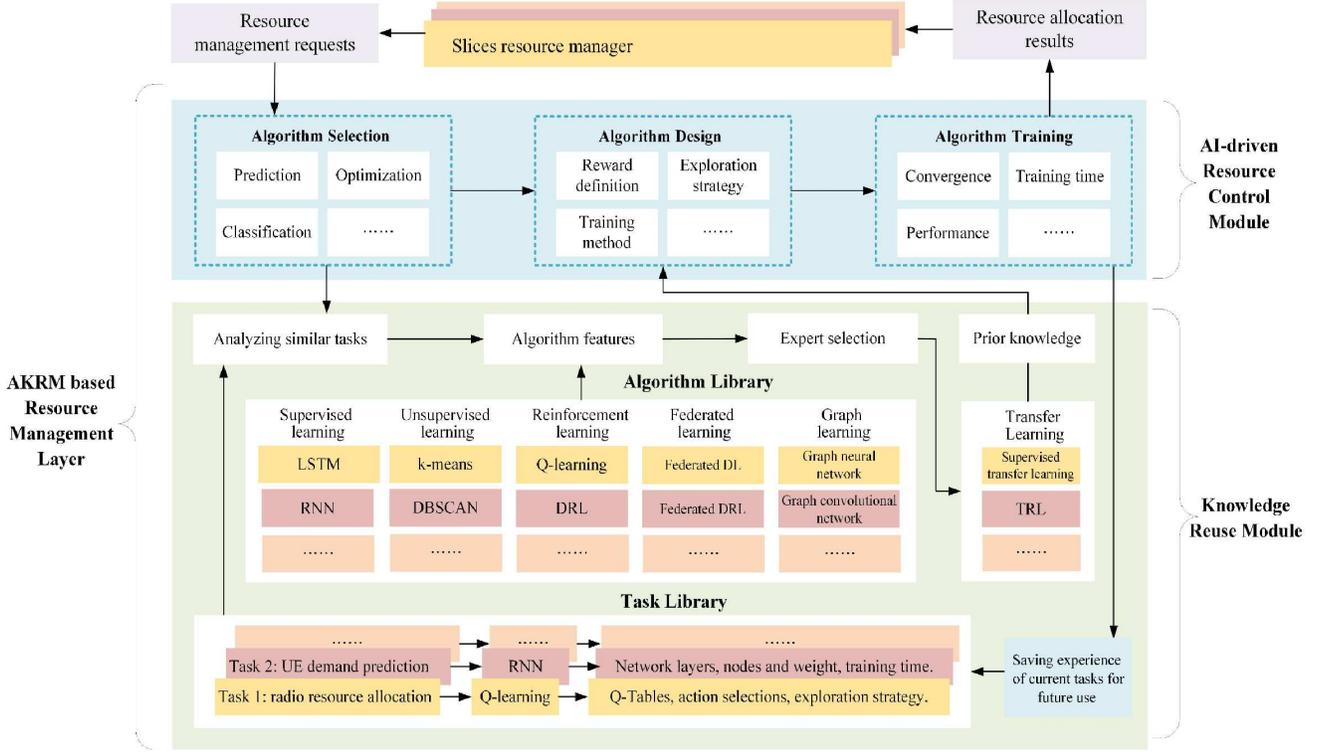}
\caption{Proposed AKRM architecture for the resource management layer.\protect\footnotemark[3]}
\label{fig2}
\vspace{-10pt}
\end{figure*}

This section introduces the AKRM-based resource management layer as Fig.\ref{fig2}, which consists of an AI-driven resource control module and a knowledge reuse module. %Combined with Fig.\ref{fig2}, here we take the resource allocation problem as an example to better explain the procedures of AKRM.

First, the slice resource managers (shown by the blocks on the top of Fig.2) will submit their requests to the management layer, and then an algorithm is selected for task processing. The task represents resource management demands from slices to optimize specific network metrics, e.g., allocating radio resources between slices to maximize the throughput and minimize the delay. Given the selected algorithm, the knowledge reuse module extracts relevant knowledge of past tasks and reuses the knowledge to improve current algorithm designs, e.g., reward function definition and exploration strategy. 

As an example, we assume the target task is Q-learning-based joint radio and computation resources allocation. Then we analyze the task similarities and algorithm features by comparing the target task with existing tasks. 
Task 1 (Q-learning-based radio resource allocation) in the task library is considered an ideal expert agent after evaluation. Consequently, the experience of task 1 can be used as prior knowledge to design a new Q-learning algorithm for joint radio and computation resource allocation. Finally, the designed Q-learning is trained, and the optimization results will be sent back to slice resource managers. On the other hand, new learning experiences generated by the current task is saved in the task library, which may be used for future tasks.

\subsection{Knowledge Reuse Module}
The knowledge reuse module is inspired by the knowledge management system (KMS), a well-known technique in both academia and industry \cite{b14}. Combined with AI techniques, here we define the knowledge reuse module to manage the knowledge generated by ML algorithms. In particular, this module captures and preserves the learning experience of AI algorithms and reuses the knowledge in the future. With transfer learning, it can extract the experience of experts and feed prior knowledge to the learner agent.  
Moreover, the new learning experience of completed tasks will be saved in the task library for future use (shown by the block at the bottom of Fig.\ref{fig2}). By updating this library, AKRM can constantly learn from the latest tasks and decisions. Unlike the algorithm-level intelligence in most existing works, the AKRM enables a system-level intelligence that learns from past experience and reuses previous knowledge on current works. In addition, each algorithm is considered a replaceable sub-module in the algorithm library, enabling higher flexibility. Finally, noting that experts are defined in various ways, including former execution experience, neighboring nodes with mature strategies, etc. The transfer learning methods can be pre-trained for each target domain, such as RL or supervised learning. 

\footnotetext[3]{The task library has two functions: 1) providing candidate experts for current tasks; and 2) saving the experience of current tasks, which may be used as new experts in the future.}

\subsection{Complexity and Compatibility Analyses}
In this section, we use MANO as an example to analyze the compatibility of AKRM with existing schemes. The MANO defined by ETSI mainly includes the network function virtualization orchestrator (NFVO), virtual network function manager (VNFM), and virtual infrastructure manager (VIM). The virtual network resources layer defined in Fig.\ref{fig1} can be deployed in the VIM, and the required resource monitoring function can be offered by the VIM. The AKRM can be included in NFVO or VNFM, depending on the required functions. In particular, the NFVO is responsible for global resource management and operation, and the algorithm library in AKRM can be customized to focus on DRL, since the global optimization may lead to a large state-action space.

Compared with the conventional one-size-fits-all scheme, the proposed AKRM scheme enables much higher flexibility for algorithm selection, design, and training. In addition, many existing techniques can be embedded to reduce the complexity. For instance, a recommender system may be deployed to select algorithms and experts, and we assume KMS is used for knowledge management in the task library module. Applying these mature techniques can significantly lower the complexity of the AKRM scheme.

\begin{figure*}[!t]
\vspace{-13pt}
\centering
\subfigure[Transfer deep reinforcement learning ]{
\includegraphics[width=14cm,height=8cm]{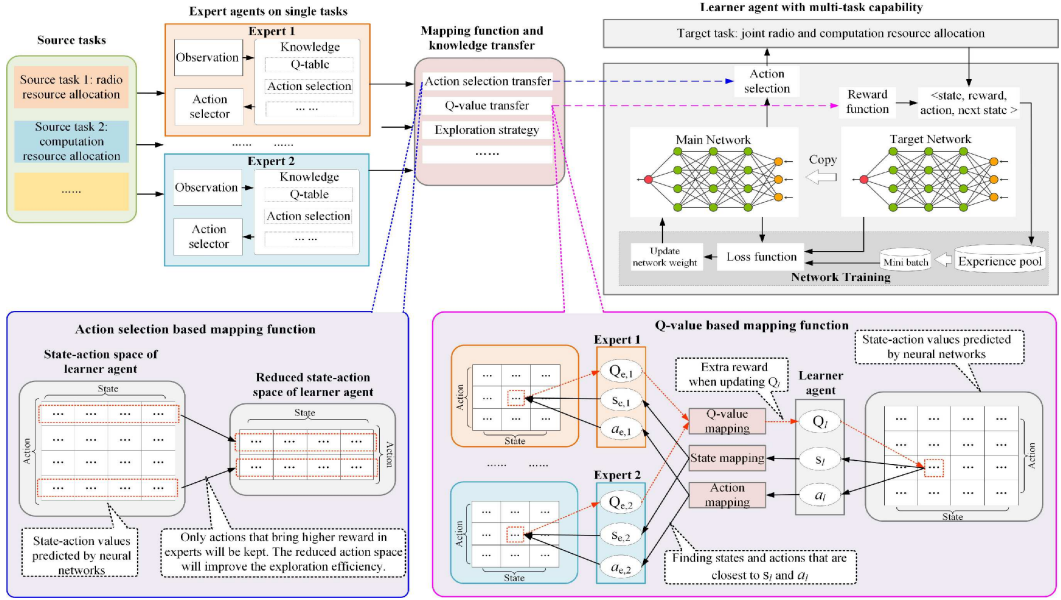}
}
\quad
\subfigure[Transfer learning-assisted neural networks]{
\includegraphics[width=14cm,height=4.2cm]{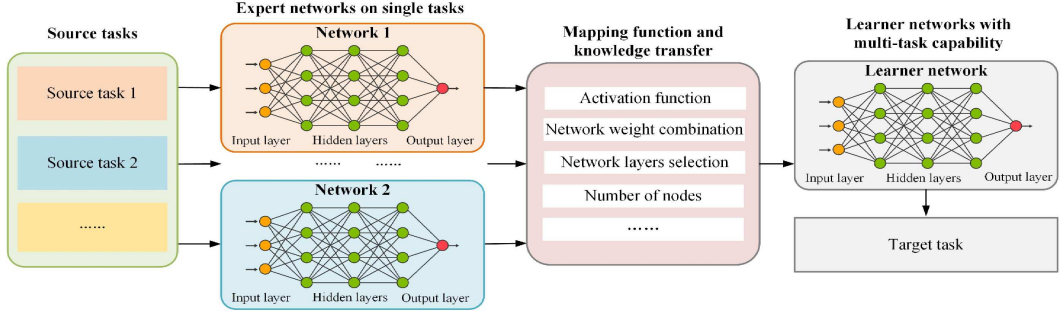}
}
\caption{Transfer learning-based knowledge reuse}
\label{fig3}
\vspace{-10pt}
\end{figure*}

\section{Transfer Learning-based Knowledge Reuse for Resource Management}
\label{s4-2}
Knowledge transfer is a critical feature of the proposed AKRM scheme. However, considering different algorithm settings such as various action and reward function definitions, prior knowledge cannot be directly used by learners. To this end, transfer learning is deployed to map the prior knowledge to current algorithms. This section will introduce two examples, namely TDRL and transfer learning-based neural networks, to better explain the knowledge transfer between different tasks.

In TDRL, the agent can utilize the knowledge from multiple experts to improve its performance on target tasks. As shown by Fig.\ref{fig3} (a), two Q-learning-based expert agents have the knowledge of radio and computation resource allocation, respectively. Then their knowledge of a single task can be used by the DRL-based learner agent for joint radio and computation resource allocation. 
Note that the prior knowledge may exist in various forms such as Q-values, action selection choices, and exploration methods, and mapping functions are required to digest the existing knowledge and feed it to learners. 
We introduce two mapping functions: action selection-based and Q-value-based methods as examples. %, shown by the bottom block of Fig.\ref{fig3} (a). 

For the action selection-based mapping function, the actions that bring higher rewards for experts will consist a new state-action space for the learner agent. The reason is that we believe optimal actions with higher rewards in experts can also bring good benefits for the learner by finding similar actions. Then a reduced action space will be applied to the learner agent, which will improve the exploration efficiency (indicated by the dark blue line in Fig.\ref{fig3} (a)). Note that the proposed mapping function can be easily applied to multi-expert scenarios by defining a weight or priority for each expert. Therefore, diverse knowledge from different experts can be comprehensively utilized by the learner. 

The Q-value-based mapping function includes Q-values as prior knowledge (shown by the pink line in Fig.\ref{fig3} (a)). To map the Q-values of expert agents to learner agent $Q_{l}(s_{l},a_{l})$, the state and action mapping functions should be first designed. The objective of the state mapping function is to find specific $s_{e,1}$ and $s_{e,2}$ that are the closest to $s_{l}$, where $s_{e,1}$ and $s_{e,2}$ are the state of expert 1 and 2, respectively. The action mapping function is defined similarly by finding the closest $a_{e,1}$ and $a_{e,2}$ for a given $a_{l}$. With state and action mapping functions, we can always observe corresponding $Q_{e,1}(s_{e,1},a_{e,1})$ and $Q_{e,2}(s_{e,2},a_{e,2})$ for any $Q_{l}(s_{l},a_{l})$. Then $Q_{e,1}(s_{e,1},a_{e,1})$ and $Q_{e,2}(s_{e,2},a_{e,2})$ can be used as extra rewards when updating $Q_{l}(s_{l},a_{l})$, which will further guide the action selection of learner. The idea is that we assume states and actions with higher Q-values in experts are very likely to bring higher rewards to learners too. With the guidance of experts, the learner is expected to achieve better performance on the target task, such as obtaining higher rewards or faster convergence.

Similarly, transfer learning can also be used for neural network training as shown by Fig.\ref{fig3} (b). The learner network may combine the weights of expert networks to initialize its weight, which will bring a jump-start for the network training. 
%In addition, the learning rate and the number of hidden layers can also be considered prior knowledge for the learner network. 
Finally, it is worth noting that the transfer learning scheme shown by Fig.\ref{fig3} (a) and (b) can be easily generalized to other AI algorithms without loss of generality. For instance, the neural network transfer scheme shown in Fig.\ref{fig3} (b) can be used by federated learning to reduce the efforts of hyperparameter tunings and network training.  

Although knowledge transfer can reduce the training complexity of learners, it still requires algorithm training experience and dataset samples of experts. The learner is expected to deploy similar model architectures with experts to achieve a smooth knowledge transfer. In addition, it is known that transfer learning is vulnerable to perturbation-based attacks caused by maliciously designed adversarial samples. Therefore, how to prevent the negative transfer or over-fitting is a critical issue, indicating that the learner may need to verify the prior knowledge before using it in practice.

\begin{figure*}[!t]
\vspace{-15pt}
\centering
\subfigure[CCDF of URLLC delay under 2 Mbps URLLC load.]{
\includegraphics[width=7.2cm,height=5.5cm]{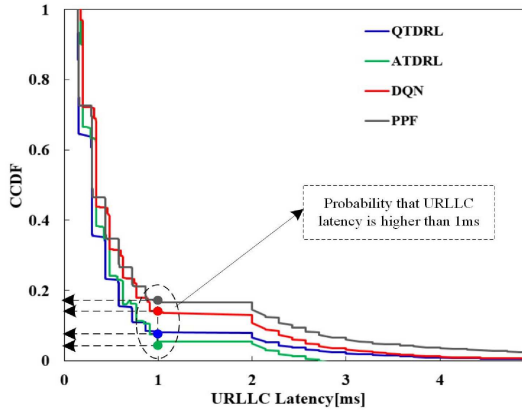}
}
\subfigure[URLLC delay and eMBB throughput various URLLC load.]{
\includegraphics[width=7.2cm,height=5.5cm]{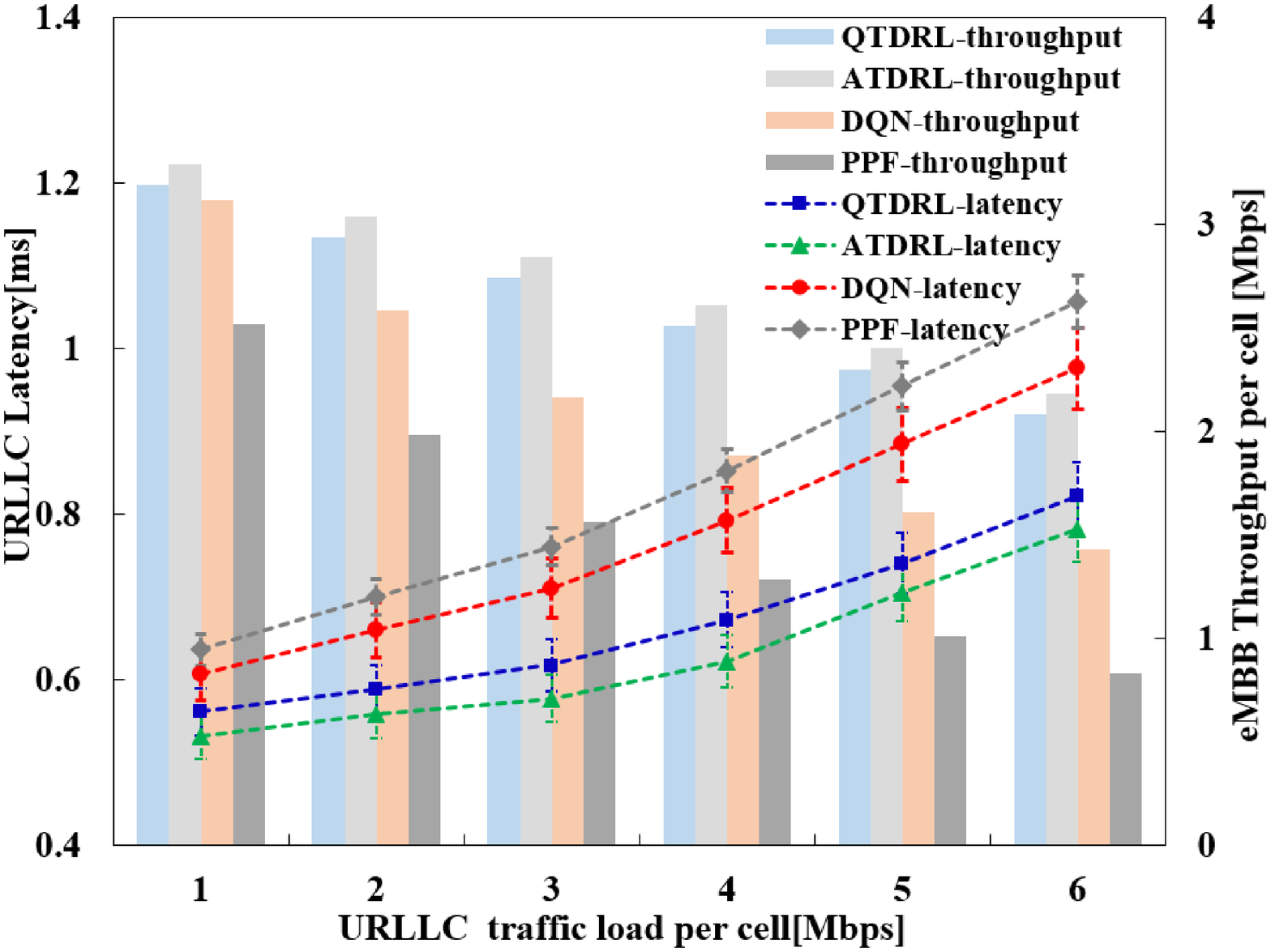}
}
\subfigure[Average URLLC delay and eMBB throughput against MEC server capacity.]{
\includegraphics[width=7.2cm,height=5.5cm]{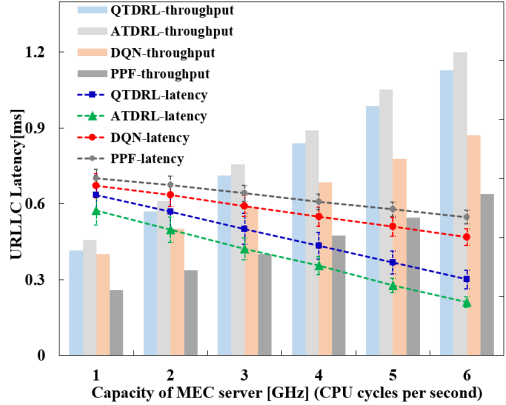}
}
\subfigure[Comparison of convergence performance.]{
\includegraphics[width=7.5cm,height=5.5cm]{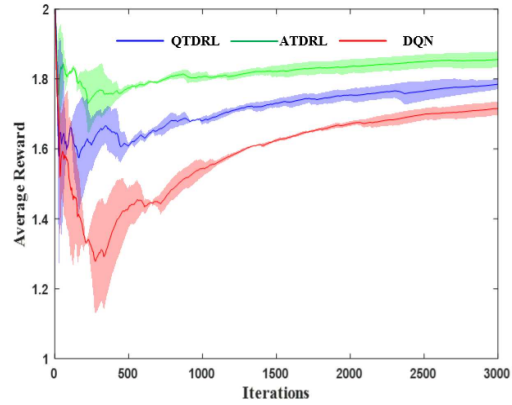}
}
\setlength{\abovecaptionskip}{0pt} 
\caption{Network delay and throughput, and convergence comparisons }
\vspace{-15pt}
\label{fig4}
\end{figure*}

\section{Transfer Reinforcement Learning-based Resource Allocation Simulation}
\label{s5}

In this section, we provide a case study of resource allocation for RAN slicing, which serves as an example of using the AI-enabled knowledge transfer scheme in 5G networks. 
Here we investigate the joint radio and computation resources allocation for RAN slicing in MEC-enabled 5G networks.
We consider two typical eMBB and URLLC slices. For radio resources, the resource blocks are considered the smallest time-frequency resources that can be allocated. For the computation resource, we assume the BSs have limited edge computing capacity (indicated by CPU cycles per second), which will be allocated between two slices to process their computation tasks. The simulation includes 3 BSs with 500 m inter-site distance, and every BS includes one URLLC slice that contains 10 UEs, and one eMBB slice that contains 5 UEs. We deploy two LSTM networks with 30 nodes as hidden layers for the main and target networks. The simulation has 3000 TTIs, including an exploration phase for the first 1000 TTIs, and an exploitation phase for the last 2000 TTIs. It is repeated for 10 runs in MATLAB with 95\% confidence intervals.

\begin{itemize}
\item \textbf{Allocation strategies:} 

i) \textbf{Q-value transfer-based deep reinforcement learning (QTDRL) for joint resource allocation :}

The QTDRL is illustrated by Fig.\ref{fig3} (a). Two Q-learning-based single-task experts have knowledge of radio and computation resource allocation, respectively. The QTDRL-based learner agent is expected to handle the joint radio and computation resource allocation based on the single-task knowledge of two experts. The learner agent can leverage the Q-values of experts as extra rewards to update its Q-values, which will guide its action selection on the target task.

ii) \textbf{Action selection transfer-based deep reinforcement learning (ATDRL) for joint resource allocation:}

Here we define another transfer learning-based method, namely ATDRL, for joint resource allocation. Compared with QTDRL, the only difference lies in the mapping function definition. The action selection-based mapping function in deployed in ATDRL for knowledge transfer (indicated by left bottom blocks in Fig.\ref{fig3} (a)). By reducing the action space, we aim to enhance the exploration efficiency of the learner.

iii) \textbf{Deep Q-learning-based joint resource allocation:}
DQN is considered as a baseline here. DQN agent has no prior knowledge about the environment, and it explores the joint radio and computation resource allocation from scratch.

iv) \textbf{Priority proportional fairness algorithm (PPF):}
PPF is a standard resource allocation method that balances the overall throughput with user fairness. PPF is considered a model-based baseline algorithm.
\end{itemize}

Fig.\ref{fig4}(a) shows the complementary cumulative distribution function (CCDF) of network delay under various traffic loads. The results show that packets in QTDRL and ATDRL have a lower probability of experiencing high delay, which is indicated by a lower CDDF curve in Fig.\ref{fig4}(a). With 2 Mbps URLLC load, the probabilities that URLLC delay is higher than 1 ms is 5.4\% and 6.8\% for QTDRL and ATDRL, while the corresponding probabilities for DQN and PPF are 12.3\% and 16.6\%, respectively. Meanwhile, the average URLLC delay and eMBB throughput per cell are given in Fig.\ref{fig4}(b), where QTDRL and ATDRL achieve lower delay for the URLLC slice and higher throughput for the eMBB slice than the two baseline algorithms.    
Moreover, we investigate the network performance under various MEC server capacities, which is given in Fig.\ref{fig4}(c). It is observed that all algorithms benefit from the increasing MEC capacity, which is indicated by lower URLLC delay and higher eMBB throughput.  
The proposed ADTRL and QDTRL still outperform model-free DQN and model-based PFF algorithms in terms of average latency and throughput.
In addition, convergence is a critical metric for ML techniques, and we compare the convergence performance in Fig.\ref{fig4}(d). The figure shows that QTDRL and ATDRL have faster convergence and higher average rewards than DQN. 

The simulations show that ATDRL and QTDRL achieve comparable network performance, but ATDRL has a better convergence, which can be explained by the reduced action space and higher exploration efficiency. In QTDRL, however, although the transferred Q-values can guide the action selection, the learner agent still needs to try a large number of action combinations, which may lower the exploration efficiency. Finally, in DQN, the agent explores the environment from scratch, and the low exploration efficiency leads to worse network performance. To summarize, transfer learning-based methods show more promising results than model-free DQN and model-based PPF algorithms, demonstrating the superiority of knowledge reuse techniques.

\section{Conclusion}
\label{s6}
AI/ML techniques offer significant opportunities for resource management of 5G and 6G networks. In this article, we have first provided background on resource management for network slicing, and surveyed related AI/ML techniques used in resource management in wireless networks. Then, we defined the AI-enabled knowledge transfer and reuse-based resource management framework, and investigated the transfer learning architectures for knowledge reuse. Our simulations showed that the proposed transfer learning-based resource allocation schemes had better network performance and significantly faster convergence than conventional deep reinforcement learning. 

%TC:ignore
\section*{Acknowledgment}
This work was supported in part by the Natural Sciences and Engineering Research Council of Canada (NSERC), the Canadian Collaborative Research and Training Experience Program (CREATE) under Grant 497981, the Canada Research Chairs Program, and the U.S. National Science Foundation under Grant CNS-2128448.

\begin{IEEEbiography}
{Hao Zhou} is a Phd candidate at the University of Ottawa. He got his B.Eng. and M.Eng degrees from Huazhong University of Science and Technology in 2016, and Tianjin University in 2019, respectively, in China. He is working towards his Phd degree at the University of Ottawa since Sep. 2019. His research interests include electric vehicles, microgrid energy trading,  resource management and network slicing in 5G. He is devoted to applying machine learning techniques for smart grid and 5G applications.     
\end{IEEEbiography}
\vspace{-40pt}
\begin{IEEEbiography}
{Melike Erol-Kantarci} is Canada Research Chair in AI-enabled Next-Generation Wireless Networks and Associate Professor at the University of Ottawa. She is the founding director of the Networked Systems and Communications Research (NETCORE) laboratory. She is also a Faculty Affiliate at the Vector Institute, Toronto. She has received numerous awards and recognitions, has delivered 70+ keynotes, tutorials and panels around the globe. She is a highly-cited prolific researcher with 200+ publications and 6 patents. She is an IEEE ComSoc Distinguished Lecturer, IEEE Senior Member and ACM Senior Member.
\end{IEEEbiography}
\vspace{-35pt}
\begin{IEEEbiography}
{H. Vincent Poor} (S’72, M’77, SM’82, F’87) is the Michael Henry Strater University Professor at Princeton University, where his interests include information theory, machine learning and network science, and their applications in wireless networks, energy systems, and related fields. He is a member of the U.S. National Academy of Engineering and the U.S. National Academy of Sciences, and a foreign member of the Chinese Academy of Sciences, the Royal Society and other national and international academies. He received the IEEE Alexander Graham Bell Medal in 2017.   
\end{IEEEbiography}
%TC:endignore


\begin{thebibliography}{00}

\bibitem{b1} A. Ksentini, and N. Nikaein, “Toward Enforcing Network Slicing on RAN: Flexibility and Resources Abstraction,” \textit{IEEE Communications Magazine}, vol. 55, no. 6, pp.102-108, Jun. 2017.

\bibitem{b2} M. Elsayed and M. Erol-Kantarci, “AI-Enabled Future Wireless Networks:
Challenges, Opportunities, and Open Issues,” \textit{IEEE Vehicular Technology Magazine}, vol. 14, no. 3, pp. 70-77, Sep. 2019. 

\bibitem{b3} ETSI, “Artificial Intelligence and future directions for ETSI (1st edition),” ETSI White Paper, Jun. 2020.

%\bibitem{b4} O-RAN Alliance, "O-RAN: towards an open and smart RAN," White paper, Oct. 2018. 

\bibitem{b4} 3GPP, “5G System; Network Data Analytics Services; Stage 3 (Release 16),” Technical specification 29.520, Aug. 2020.

%\bibitem{b5} Y. Sun, M. Peng, Y. Zhou, Y. Huang, and S. Mao, “Application of Machine Learning in Wireless Networks: Key Techniques and Open Issues,” \textit{ IEEE Communications Surveys \& Tutorials}, vol. 21, no. 4, Jun. 2019.  

%\bibitem{b7} H. Yang, X. Xie, and M. Kadoch, “Machine Learning Techniques and A Case Study for Intelligent Wireless Networks,” \textit{IEEE Network}, vol. 34, no. 3, Jan. 2020. 

\bibitem{b5} R. Shafin, L. Liu, V. Chandrasekhar, H. Chen, J. Reed, and J. Zhang, “Artificial Intelligence-Enabled Cellular Networks: A Critical Path to Beyond-5G and 6G,” \textit{ IEEE Wireless Communications}, vol. 27, no. 2, pp. 212-217, Apr. 2020. 

\bibitem{b6} C. Wang, M. D. Renzo, S. Stanczak, S. Wang, and E. G. Larsson, “Artificial Intelligence Enabled Wireless Networking for 5G and Beyond: Recent Advances and Future Challenges,” \textit{ IEEE Wireless Communications}, vol. 27, no. 1, pp. 16-23, Feb. 2020.  


\bibitem{b7} D. M. Gutierrez-Estevez, M. Gramaglia, A. D. Domenico, G. Dandachi, S. Khatibi, D. Tsolkas, I. Balan, A. Garcia-Saavedra, U. Elzur, and Y. Wang, “ Artificial Intelligence for Elastic Management and Orchestration of 5G Networks,” \textit{IEEE Wireless Communications}, vol. 26, no. 5, Oct. 2019. 

%\bibitem{b8} S. Hu, Y. Liang, Z. Xiong, and D. Niyato, “Blockchain and Artificial Intelligence for Dynamic Resource Sharing in 6G and Beyond,” \textit{IEEE Wireless Communications}, vol. 28, no. 4, Aug. 2021. 

%\bibitem{b6} A. Boudi, M. Bagaa, P. Pöyhönen, T. Taleb, and H. Flinck, “AI-Based Resource Management in Beyond 5G Cloud Native Environment,” \textit{IEEE Network}, vol. 35, no. 2, pp. 128-135, Mar. 2021. 

%\bibitem{b7} D. Bega, M. Gramaglia, R. Perez, M. Fiore, A. Banchs, and X. C. Pérez, “AI-Based Autonomous Control, Management, and Orchestration in 5G: From Standards to Algorithms,” \textit{ IEEE Network}, vol. 34, no. 6, pp. 14-20, Dec. 2020.

%\bibitem{b8} T. Taleb, I. Afolabi, K. Samdanis, and F. Z. Yousaf, “On Multi-Domain Network Slicing Orchestration Architecture and Federated Resource Control,” \textit{IEEE Network}, vol. 33, no. 5, pp. 242-252, Jul. 2019. 

\bibitem{b9} H. Zhou, and M. Erol-Kantarci, “Knowledge Transfer based Radio and Computation Resource Allocation for 5G RAN Slicing,” in \textit{Proceedings of 2022 IEEE Consumer Communications \& Networking Conference}, pp.1-6, Jan. 2022.

\bibitem{b11-1} H. Chergui, and C. Verikoukis, “Offline SLA-Constrained Deep Learning for 5G Networks Reliable and Dynamic End-to-End Slicing,” \textit{IEEE Journal on Selected Areas in Communications,} vol. 38, no. 2, pp. 350-360, Feb. 2020. 

%\bibitem{b10} M. Elsayed, M. Erol-Kantarci, and H. Yanikomeroglu, ”Transfer Reinforcement Learning for 5G New Radio mmWave Networks,” \textit{IEEE Transactions on Wireless Communications}, vol. 20, no. 5, pp. 2838-2849, May. 2021.

%\bibitem{b11} Y. Yang, F. Gao, Z. Zhong, B. Ai, and A. Alkhateeb, “Deep Transfer Learning-Based Downlink Channel Prediction for FDD Massive MIMO Systems,” \textit{IEEE Transactions on Communications}, vol. 68, no. 12, pp. 7485-7497, Dec. 2020.  

%\bibitem{b12} H. H. Yang, Z. Liu, T. Q. S. Quek, and H. V. Poor, “Scheduling Policies for Federated Learning in Wireless Networks,” \textit{IEEE Transactions on Communications}, vol. 68, no. 1, pp. 317-333, Jan. 2020.

\bibitem{b12} S. Messaoud, A. Bradai, O. B. Ahmed, P. T. A. Quang, M. Atri, and M. S. Hossain, “Deep Federated Q-Learning-Based Network Slicing for Industrial IoT,” \textit{IEEE Transactions on Industrial Informatics}, vol. 17, no. 8, pp. 5572-5582, Aug. 2021.

\bibitem{b13} H. Wang, Y. Wu, G. Min, and W. Miao, “A Graph Neural Network-based Digital Twin for Network Slicing Management,” \textit{IEEE Transactions on Industrial Informatics}, vol. 18, no. 2,  pp. 1367-1376, Feb. 2022.

\bibitem{b10} T. Mai, H. Yao, N. Zhang, W. He, D. Guo, and M. Guizani, “Transfer Reinforcement Learning Aided Distributed Network Slicing Optimization in Industrial IoT,” \textit{IEEE Transaction on Industrial Informatics}, vol. 18, no. 6, pp. 4308-4316, Jun. 2022.

\bibitem{b14} M. F. Manesh, M. M. Pellegrini, G. Marzi, and M. Dabic, “Knowledge Management in the Fourth Industrial Revolution: Mapping the Literature and Scoping Future Avenues,” \textit{IEEE Transactions on Engineering Management,} vol. 68, no. 1, pp. 289-300, Feb. 2021.

\end{thebibliography}
\end{document}